\documentstyle[12pt,twoside,psfig]{article}

\def\be{\begin{equation}}
\def\ee{\end{equation}}
\def\bea{\begin{eqnarray}}
\def\eea{\end{eqnarray}}
\def\psibar{\overline{\psi}}
\def\diag{{\rm diag}}
\def\vac{|0\rangle }
\def\lvac{\langle 0| }
\def\calD{{\cal D}}
\def\calB{{\cal B}}
\def\nn{\nonumber}
\def\calU{{\cal U}}

\def\d{\partial}

\def\g{\gamma}

\def\m{\mu}
\def\n{\nu}
\def\o{\omega}
\def\s{\sigma}                    
\def\t{\tau}

\def\O{\Omega}

\def\M{{\rm M}}
\def\E{{\rm E}}
\def\dslash{\partial\!\!\! /}
\def\kslash{k\!\!\! /}

\def\inv{^{-1}}
\def\wk{\o_{\vec{k}}}
\def\t{\tau}
\def\psidag{\psi^\dagger}
\newcommand{\rf}[1]{~(\ref{#1})}

\begin{document}

\thispagestyle{empty}
\begin{flushright}
{\sc ITP-SB}-97-7
\end{flushright}
\vspace{1cm}
\setcounter{footnote}{0}
\begin{center}
{\LARGE{A Wick Rotation for Spinor Fields: the Canonical
         Approach}  
    }\\[14mm]

\sc{
Andrew Waldron\footnote{e-mail: 
wally@insti.physics.sunysb.edu},}\\[5mm]
{\it Institute for Theoretical Physics\\
State University of New York at Stony Brook\\
Stony Brook, NY 11794-3840, USA}\\[20mm]

{\sc Abstract}\\[2mm]
\end{center}
Recently we proposed a new Wick rotation for Dirac spinors
which resulted in a hermitean action in Euclidean space.
Our work was in a path integral context, however, in this
note, we provide the canonical formulation of the
new Wick rotation along the lines of the proposal of 
Osterwalder and Schrader. 

\vfill

\newpage

\section{Introduction and Review.}

Field theories Wick rotated to Euclidean space are the subject
of an enormous body of research.
In particular, 
modern non-perturbative studies of supersymmetric theories
(for example instantons and the study of Donaldson
invariants for compact manifolds) depend on the introduction
of Euclidean field theories. Clearly then, it is of crucial
importance to understand how one performs a Wick rotation for
spinors. 

In a previous publication in this journal~\cite{us} 
we observed that 
there existed two apparently distinct approaches to Dirac
spinors in Euclidean space. Namely the approach of Osterwalder
and Schrader~\cite{OS2} (OS) in which the  fields $\psi$ and its
Dirac conjugate $\psibar$ are taken to be independent and
hermiticity is forsaken and the approach of
Schwinger~\cite{Schwinger} and later Zumino~\cite{Zumino}
in which spinor degrees of freedom are undoubled and the action
in Euclidean space is hermitean. Within a path integral context,
the distinction between integrating over fields $\psi$ and
$\psi^\dagger$ versus independent fields $\psi$ and 
$\chi^\dagger$ ($\psi\neq\chi$)
is only semantic due to the algebraic nature of Grassmann
integration, so the real puzzle was therefore to understand
how Schwinger was able to maintain hermiticity whereas
OS did not. This problem was solved by introducing a new Wick
rotation for Dirac spinors which acted only on the fundamental
fields and coordinates.

For vectors $A_\mu(t,\vec{x})$, the Wick rotation to Euclidean 
space is performed by
transforming both the time coordinate 
$t\rightarrow-i\tau$ and vector indices by a matrix
${\Omega_\mu}^\nu=\diag(i,1,1,1)$,
i.e., $A_\mu(t,\vec{x})\rightarrow{\Omega_\mu}^\nu
A_\mu^\E(\tau,\vec{x})$. However for complex vectors 
the complex conjugate $A_\mu^\dagger$ transforms under the same
matrix ${\Omega_\mu}^\nu$ rather than
${{\Omega_\mu}^\nu}^\dagger$.
This observation led us to introduce the 
following Wick rotation for Dirac spinors~\cite{us}: 
\bea
\psi(t,\vec{x})&\rightarrow&S(\theta)
\psi_\theta(t_\theta,\vec{x})\\
\psi^\dagger(t,\vec{x})
&\rightarrow&\psi_\theta^\dagger(t_\theta,\vec{x})S(\theta)\\
t&\rightarrow&e^{-i\theta}t_\theta,
\eea
where\footnote{Our conventions are as follows, the Minkowski
Dirac matrices $\g^\m=\big(\g^0\equiv-i\g^4,\vec{\g}\big)^\m$, 
where
$\g^4$ and $\vec{\g}$ are hermitean and
$\{\g^\m,\g^\n\}=2\eta^{\m\n}=2\diag(-1,1,1,1)^{\m\n}$. The
matrix $\g^5\equiv\g^1\g^2\g^3\g^4$ is hermitean.}
\be
S(\theta)=e^{\g^4\g^5\theta/2}, \ \ \theta\in[0,\pi/2].
\ee
The matrix $S(\theta)$ is unitary and the parameter $\theta$ is
introduced to provide a continuous\footnote{Note that Lorentz
invariance of the intermediate interpolating theories is
obtained only if one complexifies so that $\psi_\theta$ and
$\psi_\theta^\dagger$ are independent spinors and hermiticity of
the interpolating Dirac action is lost. Our claim is not that
$\psi$ and $\psi^\dagger$ remain dependent under Wick rotation,
but rather, that for Dirac spinors, hermiticity at the endpoint
(the Euclidean theory) may be regained. In this sense one may
think of hermiticity as a symmetry property of the Euclidean
theory.} interpolation between
the Minkowski and Euclidean theories.
At the initial value $\theta=0$, $S(\theta=0)=I$ and
$\psi_{\theta=0}\equiv\psi$,
$\psi^\dagger_{\theta=0}\equiv\psi^\dagger$
and $t_{\theta=0}\equiv t\equiv x^0\equiv -x_0$ 
take their usual Minkowski values,
whereas at the endpoint $\theta=\pi/2$,
$S(\theta=\pi/2)=e^{\g^4\g^5\pi/4}\equiv S$ 
and
$\psi_{\theta=\pi/2}\equiv\psi_\E$,
$\psi^\dagger_{\theta=\pi/2}\equiv\psi_\E^\dagger$
and $t_{\theta=\pi/2}\equiv \tau\equiv x^4\equiv x_4$ 
are their Euclidean counterparts. Observe that $S(\theta)\g^4$ 
$=$ $\g^4 S\inv(\theta)$ whereby
\be
\psibar(t,\vec{x})\equiv\psi^\dagger(t,\vec{x})\g^4
\rightarrow \psi^\dagger_\theta
(t_\theta,\vec{x})\g^4 S\inv(\theta),
\ee
so that our Wick rotation induces a similarity transformation
upon the Dirac matrices in Dirac bilinears $\psibar\,
\Gamma_A \psi$ for some combination of Dirac matrices
$\Gamma_A$, 
\bea
\vec{\g}(\theta)&\equiv&S\inv(\theta)\vec{\g}S(\theta)=\vec{\g}
\equiv\vec{\g}_\E\\
\g^4(\theta)&\equiv&S\inv(\theta)\g^4S(\theta)=\g^4\cos\theta
+\g^5\sin\theta\\
\g^5(\theta)&\equiv&S\inv(\theta)\g^5S(\theta)=-\g^4\sin\theta
+\g^5\cos\theta.
\eea
Note then that $\g_{\theta=\pi/2}^4\equiv\g_\E^4=\g^5$,
$\g_{\theta=\pi/2}^5\equiv\g_\E^5=-\g^4$
whereby the Wick rotation of the Dirac conjugate spinor
$\psibar$ yields
\be
\psibar(t,\vec{x})=\psi^\dagger(t,\vec{x})\g^4
\rightarrow
-\psi^\dagger_\E(\tau,\vec{x})\g_\E^5e^{-\g^4\g^5\theta/4}
\ee
and the troublesome $\g^4$ in $\psibar$ has been reinterpreted as
$-\g^5_\E$ in the Euclidean theory.
Applying this Wick rotation to the action for a free massive
Dirac spinor
\be
\frac{i}{\hbar}S_\M\equiv
-\frac{1}{\hbar}\int
dtd\vec{x}\psi^\dagger(t,\vec{x})\g^4(\g^0\frac{\d}{\d
t}+\vec{\g}.\vec{\d}+m)\psi(t,\vec{x}),
\ee
we obtain 
\be
-\frac{1}{\hbar}S_\E\equiv\frac{1}{\hbar}\int d\tau 
d\vec{x}\psi^\dagger_\E(\tau,\vec{x})
\g^5_\E(\g^4\frac{\d}{\d\tau}+\vec{\g}.\vec{\d}+m)
\psi(\tau,\vec{x}).
\ee
This action is hermitean, $SO(4)$ invariant and is the
result of a Wick rotation acting as an analytic continuation
$t\rightarrow-i\tau$ and a {\it simultaneous rotation on
spinor indices}.

The above Wick rotation produces a Euclidean theory for 
spinor fields whose action, in the exponent of a Euclidean path
integral, yields Euclidean Greens functions which are related
to the usual Minkowski Greens functions by analytic continuation
and a rotation on spinor indices by the matrix $S$ introduced
above.    
However, along with necessary and sufficient conditions
for a Euclidean theory to produce the analytically continued
counterparts of the Greens function of a given Minkowski
theory~\cite{OS1}, 
Osterwalder and Schrader~\cite{OS2} have explicitly constructed a
canonical Euclidean theory in terms of Euclidean Dirac 
spinor fields
acting in a Euclidean Fock space whose Greens
functions are the analytic continuations of the corresponding 
Minkowski Greens functions. In the remainder of this note, 
we shall reformulate their work in the context of our
Wick rotation for Dirac spinors.

\section{The Main Ingredients of the OS construction.}

We begin by briefly sketching the main ingredients of the
OS construction. Firstly we analytically continue canonical
Minkowski fields to imaginary times $t\rightarrow-i\tau$.
The canonical (Heisenberg) 
Minkowski field satisfying the free Dirac 
equation $(\dslash+m)\psi=0$ is given by
\be
\psi(t,\vec{x})=e^{iH_0t}\psi(0,\vec{x})e^{-iH_0t},
\label{mf}
\ee
with the usual mode expansion
\be
\psi(0,\vec{x})=\int\frac{d\vec{k}}{(2\pi)^{3/2}}\left(
b_{\vec{k}}\cdot u_{\vec{k}}e^{i\vec{k}.\vec{x}}+
d^\dagger_{\vec{k}}\cdot v_{-\vec{k}}e^{-i\vec{k}.\vec{x}}
\right)
\ee
and normal ordered free Hamiltonian 
$H_0=\int d\vec{k}\o_{\vec{k}}(b^\dagger_{\vec{k}}\cdot
b_{\vec{k}}+d^\dagger_{\vec{k}}\cdot d_{\vec{k}})$ where
$\o_{\vec{k}}=(\vec{k}^2+m^2)^{1/2}$. We denote the sum over
spin polarizations by a dot, i.e.,
$b_{\vec{k}}\cdot u_{\vec{k}}=
\sum_{r=1,2}b_{\vec{k}}^r\cdot u_{\vec{k}}^r$.
The orthonormal 
spinor wave functions satisfy $(i\kslash+m)u_{\vec{k}}=0=
(-i\kslash+m)v_{-\vec{k}}$ and spin polarization sums
\be
u_{\vec{k}}\cdot\overline{u}_{\vec{k}}=
\frac{-i\kslash+m}{2\wk}\ ; \
v_{-\vec{k}}\cdot\overline{v}_{-\vec{k}}=
\frac{-i\kslash-m}{2\wk}.
\ee
Defining the usual Minkowski vacuum 
$\vac$ via $d_{\vec{k}}\vac=0=b_{\vec{k}}\vac$
and imposing commutation relations for the modes
\be
\{b_{\vec{k}},b^\dagger_{\vec{k}^\prime}\}\
=\delta^3(\vec{k}-\vec{k}^\prime){\sc 1}=
\{d_{\vec{k}},d^\dagger_{\vec{k}^\prime}\}
\ee
(we have suppressed the polarization indices $r,s=\pm$ so
${\sc 1}$
denotes $\delta^{rs}$) one obtains the two-point function
(propagator)
\be
\lvac T \psi(x)\psibar(y)\vac=-i\int\frac{dk_0d^3k}{(2\pi)^4}
\frac{-i\kslash+m}{k^2+m^2-i\epsilon}e^{ik_\m(x-y)^\m}
\equiv\calD(x-y).
\label{prop}
\ee
The analytically 
continued Minkowski fields are constructed by
allowing the Minkowski field~(\ref{mf}) to undergo 
imaginary time evolution with
$t=-i\tau$,
\bea
\psi(-i\t,\vec{x})&=&e^{H_0\t}\psi(0,\vec{x})
e^{-H_0\t}\nonumber\\&=&
\int\frac{d^3k}{\sqrt{(2\pi)^3}}\left\{
b_{\vec{k}}\cdot u_{\vec{k}}e^{-\wk\tau+i\vec{k}.\vec{x}}
+d^\dagger_{\vec{k}}\cdot v_{-\vec{k}}
e^{\wk\tau-i\vec{k}.\vec{x}}
\right\}
\label{minpsi}
\eea
The continuation of the Dirac conjugate field $\psibar$ 
is defined in the same way
\bea
\psibar(-i\t,\vec{x})&=&e^{H_0\t}
\psidag(0,\vec{x})i\g^0e^{-H_0\t}
\nonumber\\&=&
\int\frac{d^3k}{\sqrt{(2\pi)^3}}\left\{
b_{\vec{k}}^\dagger\cdot 
\overline{u}_{\vec{k}}
e^{+\wk\tau-i\vec{k}.\vec{x}}
+d_{\vec{k}}\cdot 
\overline{v}_{-\vec{k}}e^{-\wk\tau+i\vec{k}.\vec{x}}
\right\}\nonumber\\&=&
\left[\psi(\vec{x},+i\tau)\right]^\dagger i\g^0.
\label{psib}
\eea
It is important to realize from the last equality in\rf{psib}
that hermitean conjugation and analytic continuation do not
commute, but rather an additional ``Euclidean time reversal''
$\t\rightarrow-\t$ is required. The concept of reflection
positivity follows from this remark.
The continued free two-point function, or propagator, can now 
straightforwardly be 
constructed
from the continued Minkowski fields,
\bea\hspace{-.6cm}
\calD(-i\t+i\sigma,\vec{x}-\vec{y})&\equiv &
\lvac
\widetilde{T}\psi(-i\t,\vec{x})\psibar(-i\s,\vec{y})
\vac\nonumber\\
&=&\theta(\t-\s)\lvac \psi(-i\t,\vec{x})\psibar(-i\s,\vec{y})\vac
\nonumber\\&&
-\theta(\s-\t)\lvac
\psibar(-i\s,\vec{y})\psi(-i\t,\vec{x})\vac\nonumber\\
&=&\int\frac{dk_4d^3k}{(2\pi)^4}\frac{-i\kslash^{\rm OS}
+m}{k^2_\E+m^2}
e^{ik^\E_\m(x-y)^\E_\m}\equiv\calD_\E(x^\E-y^\E).\label{cont}
\eea
Let us make a few comments. 
The symbol $\widetilde{T}$ denotes time ordering,
but now with respect to $\t$, the continued time. Further,
$\kslash^{\rm OS}\equiv\g^4k_4+\vec{\g}.\vec{k}$ so that 
{\it spinor indices undergo no rotation} in the OS 
approach,
rather an $i$ is simply ``borrowed'' from the relation 
$\g^0=-i\g^4$. Also $k^\E_\m x^\E_\m\equiv k_4\t+\vec{k}.\vec{x}$
is the Euclidean inner product.
Furthermore, it is easy to show that this result for
$\calD(-i\t,\vec{x})$ is exactly the same as that
obtained from the function $\calD(t,\vec{x})$ in~(\ref{prop})
by a direct (unique) analytic continuation in the time variable.

So far we have done nothing except consider 
the usual Minkowski fields
acting in the Minkowski Fock space 
but at imaginary values of the time
coordinate. However in the last line 
of~(\ref{cont}) we denoted 
$\calD(-i\t,\vec{x})=\calD_\E(x^\E)$ 
because the next step is to construct
Euclidean fields acting in a 
Euclidean Fock space
whose two-point function is $\calD_\E(x^\E)$. 
General Greens functions can be reconstructed
via use of the Wick theorem.

The time coordinate should play
no preferred r\^{o}le in Euclidean
space so the OS fields anticommute
at all points $x^\E$ and $y^\E$ ($x^\E$ and $y^\E$
are now, of course, four dimensional). 
Hence there is no time ordering
in their Euclidean formulation and 
the Euclidean fields are expanded 
in terms of modes depending on Euclidean
four-momenta. 
In any case there are no plane wave solutions to
$(-\Box_\E+m^2)\psi_\E=0$ in flat Euclidean space,
so that, in the OS construction, Euclidean fields are off-shell.
Hence their Euclidean fields have a form which is rather close to
that of a five-dimensional Minkowski field at 
zero time. Explicitly
the Euclidean canonical Dirac spinor of OS is given by
\be
\psi_\E(x)=\int\frac{d^4k}{\sqrt{(2\pi)^4 \O_k}}\left\{
B_k\circ U_ke^{ikx}+D_k^\dagger\circ V_{-k}e^{-ikx}\right\},
\label{psiexp}
\ee 
where $\O_k\equiv(k^2+m^2)^{1/2}$,
$e^{ikx}=e^{i(k_4\t+\vec{k}.\vec{x})}$ and from now one we drop
the subscript $_\E$ on the Euclidean four-vectors $k$ and $x$. 
The mode operators $B_k$ and $D^\dagger_k$ act in a 
Euclidean Fock
space with vacuum $\vac_\E$ defined such that
$B_k\vac_\E=0=D_k\vac_\E$. The only non-vanishing 
anticommutation relations of the modes
now give a four dimensional delta function,
\be
\{B_k,B^\dagger_{k^\prime}\}=\delta^4(k-k^\prime){\sc 1}=
\{D_k,D_{k^\prime}^\dagger\},\label{ac}
\ee 
where we again suppress polarization indices 
so that $\sc{1}$ denotes 
$\delta^{RS}$. However, let us stress that the indices 
$R$ and $S$ no longer run
over values $1,2$, rather it is necessary to 
{\it double} the
spin polarization degrees of freedom whereby $R,S=1,..,4$.
and we denote
$B_k\circ U_k\equiv\sum_{R=1}^{4}B_k^R U_k^R$. 
The spinor wave 
functions $U_k$ and $V_{-k}$ do not satisfy any 
equations of motion. 
At this point the fields $\psi_\E(x)$ in\rf{psiexp} and 
$\psi(-i\t,\vec{x})$ in\rf{minpsi} are 
totally unrelated, they act in different Fock spaces.

The next task is to construct a
conjugate momentum field 
to $\psi_\E(x)$ (i.e., the analogue of $\psi^\dagger$ in 
the Minkowski case). 
In the OS proposal, 
the answer is no longer $[\psi_\E(x)]^\dagger$
(although a key feature of the canonical formulation of our new
Wick rotation is that this this property will be retained),
instead the operation of hermitean conjugation 
is replaced by the composition of hermitean conjugation and a 
unitary involution $\Theta$ defined as follows.
For Bose fields $\phi_\E(\tau,\vec{x})$
(which may be easily be treated in the OS approach
without doubling, see~\cite{OS2} for details) $\Theta$ 
acts simply as
Euclidean time reversal,
\be
\Theta\phi_\E(\tau,\vec{x})\Theta\inv=\phi_\E(-\tau,\vec{x})
=\phi_\E(\Theta x),\label{blizzard}
\ee
where $\Theta (\t,\vec{x})=(-\t,\vec{x})$.
This definition is motivated by the remark 
above that for continued Minkowski 
fields, 
one needed an additional Euclidean time reversal when comparing
imaginary time evolution of the hermitean conjugated field 
with the 
hermitean conjugate 
of the imaginary time evolved field (see~(\ref{psib})).
For spinors however, the action of $\Theta$ is more subtle.
Define a field $\chi_\E^\dagger(x)$, the Euclidean 
analogue of the Minkowski field
$\psibar(\vec{x},t)$, by
\be
\chi^\dagger_\E(x)=\int\frac{d^4k}{\sqrt{(2\pi)^4 \O_k}}\left\{
B^\dagger_k\circ W^\dagger_k e^{-ikx}+D_k\circ 
X_{-k}^\dagger e^{ikx}\right\}.
\label{chi}
\ee
Since we already doubled
the number of spin
polarizations ($R,S=1,..4$),
it would be a redoubling if the field $\chi^\dagger_\E(x)$
were independent\footnote{Contrast this to an OS path
integral approach for Dirac spinors in which the ``doubling'' of
spinor degrees of freedom in Euclidean space is introduced by
taking the field $\chi_\E^\dagger$ to be independent of
$\psi_\E$.} of $\psi_\E(x)$. Rather,
$\chi^\dagger_\E$ is related to $\psi_\E$ by both
hermitean conjugation and the action of $\Theta$,
\be
\chi^\dagger_\E(\tau,\vec{x})
\equiv\Theta\inv\psi_\E^\dagger(-\tau,\vec{x})i\g^0
\Theta=\Theta\inv
\psi_\E^\dagger(\Theta x)i\g^0\Theta. \label{theta}
\ee
The relation~(\ref{theta})
together with the expansions~(\ref{chi}) and~(\ref{psiexp}) 
imply that $\Theta$ has 
a more complicated action on the modes (and in turn states
in the Euclidean Fock space) such that
\be
\Theta\inv B^\dagger_{\Theta k} 
\circ U_{\Theta k}^\dagger i\g^0 \Theta=
B^\dagger_k\circ W^\dagger_k\ ; \ 
\Theta\inv D_{\Theta k} \circ V_{-\Theta k}^\dagger i\g^0\Theta
=D_k\circ X_{-k}^\dagger
\ee
where $\Theta k = (-k^4,\vec{k})$. Given the explicit forms 
(see~\cite{OS2}) of the 
spinor wavefunctions $U_k$, $V_{-k}$, $W_k$ and $X_{-k}$ 
one can write 
down the action of $\Theta$ on the 
mode operators $B_k$ and $D_k$.
For our purposes 
it is enough to note that the spinor wavefunctions
satisfy spin polarization sums constructed such
that one obtains the
desired two-point function in\rf{cont},
\be
U_{k}\circ W_{k}^\dagger=\frac{-i\kslash^{\rm OS}+m}{\O_k}\ ; \
V_{-k}\circ X_{-k}^\dagger=\frac{-i\kslash^{\rm OS}-m}{\O_k}.
\label{sps}
\ee
Defining $\Theta\vac_\E=\vac_\E$ one can then also 
calculate the action of
$\Theta$ on states in the Euclidean Fock space. The mode 
expansions~(\ref{psiexp}) and~(\ref{chi}) along with the spin 
polarization sums in~(\ref{sps}) yield
\be
\{\psi_\E(x),\chi^\dagger_\E(y)\}=0.
\label{acc}
\ee
Furthermore, using~(\ref{psiexp}), 
(\ref{ac}), (\ref{chi}) and~(\ref{sps}) it is easy to verify the
following equalities
\be
\calD_\E(x-y)\equiv{}_\E\!
\lvac\psi_\E(x)\chi_\E^\dagger(y)\vac_\E
=-_\E\!\lvac\chi_\E^\dagger(y)\psi_\E(x)\vac_\E
=\calD(-i\t+i\s,\vec{x}-\vec{y}),
\label{skunk}
\ee
where $D_\E(x-y)$ in~(\ref{skunk}) agrees with  $D_\E(x-y)$ 
in~(\ref{cont}) so that the Euclidean two-point
function without time ordering reproduces the continued, time
ordered Minkowski two-point function.

The final ingredient is the relation between states in the 
Euclidean Fock space, and those in the physical Minkowski
Hilbert space. This is provided by the following mapping 
\be
W:|X\rangle_\E\rightarrow |WX\rangle_\M,
\ee
from an arbitrary state $|X\rangle_\E$ in the Euclidean Fock
space to some state $|WX\rangle_\M$ in the Minkowski
Hilbert space. We shall call this mapping the ``OS--Wick map''
and it is defined as follows. 
A general state in the Euclidean Fock space~\cite{OS2}
may be represented as
\be 
|X\rangle_\E=\int d^4x_1...d^4x_{m+n}f_1(x_1)...f_{m+n}(x_{m+n})
:\underbrace{\psi_\E(x_1)...\chi_\E^\dagger(x_{m+n})}_{\mbox{$m$ 
$\psi_\E$'s, $n$  $\chi^\dagger_\E$'s}}
:\vac_\E.\label{state}
\ee
The functions $f_i(x_i)$ are some choice of test functions 
and
it is convenient to normal order 
this expression as denoted by $:\ :$
by which we mean all annihilation 
operators $B_k$ and $D_k$ are to 
be pulled to the right.
The OS--Wick map may now be defined by its action on the state
$|X\rangle_\E$ in\rf{state},
\be
W|X\rangle_\E=|WX\rangle_\M\equiv 
:\psi(-i\t_1,\vec{x}_1)...\psibar(-i\t_{m+n},\vec{x}_{m+n}):\vac.
\label{X}
\ee
where, for brevity, we have suppressed the smearing by
test functions
$f_i(x_i)$. The fields 
$\psi(-i\t_1,\vec{x}_1)$ and
$\psibar(-i\t_{m+n},\vec{x}_{m+n})$ are precisely the continued
Min\-kowski fields defined in\rf{minpsi} and\rf{psib}, 
respectively, above.
In~\cite{OS2}, the following central theorem is proven
\be
_\E\!\langle\Theta X|Y\rangle_\E=
\ _\M\!\langle WX|WY\rangle_\M.\label{mr}
\ee
which states that inner products of states in the 
Euclidean Fock space are related to those in the 
Minkowski Hilbert space by the OS--Wick map and the unitary
involution $\Theta$. 
The inner product in the Minkowski
Hilbert space should be positive definite, whereby we
immediately obtain the OS reflection positivity condition
\be
_\E\!\langle \Theta X|X\rangle_\E\geq 0.
\label{positive}
\ee
As yet we have made no mention of how dynamics are included 
in this proposal, but at this point we refer the reader to 
the original work of Osterwalder and Schrader~\cite{OS2}.
Let us now give the generalization of the above construction to
include our new Wick rotation.

\section{The Canonical Formulation of the New Wick Rotation.}

In~\cite{OS2}, it is argued that the 
field $\chi^\dagger_\E$ cannot be replaced 
by $\psi^\dagger_\E$ 
since the two-point function,
\be
_\E\!\lvac\psi_\E(x)\chi_\E^\dagger(y)\vac_\E=
\int\frac{d^4k}{(2\pi)^4}\frac{-i\kslash^{\rm OS}+m}{k^2_\E+m^2}
e^{ik_\E(x-y)_\E},\label{schnoot}
\ee
is then inconsistent because
only the left hand side is 
invariant under hermitean conjugation and 
the interchange of $x$ and $y$.
In light of our new Wick rotation, the remedy is obvious. 
One should replace $\kslash^{\rm OS}$ by
$\kslash^\E=k^\m_\E\g^\m_\E$ where the  
matrices $\g^\m_\E$ are
defined in the introduction (and were obtained via a
similarity transformation induced by the rotation of spinor
indices) and put 
\be
\chi_\E^\dagger=-\psi_\E^\dagger\g_\E^5.
\ee
In order to incorporate our rotation of spinor indices in
the OS proposal,
we replace the OS--Wick  map $W$
by a new OS--Wick map $\widetilde{W}$, defined as follows
\be
\widetilde{W}:|X\rangle\rightarrow|\widetilde{W}X\rangle_\M
\ee
where, in general,
\be
|X\rangle_\E=:\psi_\E(x_1)...\psi_\E^\dagger(x_{n+m})(-\g_\E^5):
|0\rangle_\E
\ee
and
\be
|\widetilde{W}X\rangle_\M=:S\psi(-i\t_1,\vec{x}_1)...
\psibar(-i\t_{n+m},\vec{x}_{n+m})S^{-1}:|0\rangle_\M
\label{hendrix}
\ee
with $S=e^{\g^4\g^5\pi/4}$. Hence, Euclidean Greens functions
are related to their continued Minkowski Greens functions 
by an additional rotation of spinor indices. For example, for
the free two-point function, we find
\bea
_\E\!\lvac\psi_\E(x)\psi_\E^\dagger(y)(-\g_\E^5)\vac_\E&=&
\int\frac{d^4k}{(2\pi)^4}\frac{-i\kslash^\E+m}{k^2+m^2}
e^{ik(x-y)}\nn\\&=&
S\inv \calD(-i\t+i\s,\vec{x}-\vec{y})S.
\label{2}
\eea
We must now construct the field $\psi_\E$, the Euclidean Fock
space and the unitary involution $\Theta$. Let us briefly 
mention two unfruitful avenues before we give our solution.
The first would be to require $W_k^\dagger$ in\rf{chi}
and $U_k$ in\rf{psiexp} to satisfy
\be
W_k^\dagger=-U_k^\dagger\g_\E^5.
\label{gibson}
\ee
(along with an analagous condition on $V_{-k}$ and $X_{-k}$).
However, we must still reproduce the correct continued Minkowski
two-point function which requires
$U_k\circ W_k^\dagger=-i\kslash^\E+m$ for which there is no
solution when\rf{gibson} holds\footnote{To see this, take a
basis in which $\g_\E^5={\rm
diag}(1,1,-1,-1)$ and the other Dirac matrices are
off-diagonal. Multiplying by $-\g_\E^5$ and tracing yields
$U_k^R\equiv0$ $(R=1,...,4)$.}.

A second, bolder proposal would be to notice that the 
expressions\rf{psiexp} and\rf{chi} are reminiscent of those of a 
five (5=1+4) dimensional Dirac spinor at time $t^5=0$, except
that the polarization sums denoted by ``$\circ$'', 
should run over
two, instead of four values.  
The necessity to reproduce the continued $3+1$
dimensional
Minkowski propagator,
however, can be used to rule out directly replacing $\psi_\E$
and $\chi_\E^\dagger$ by their zero time, five dimensional
counterparts.

In this light, we consider the more general ansatz
\be
\psi_\E(x)=\int \frac{d^4k}{\sqrt{(2\pi)^4}}\left\{
\calB_ke^{ikx}+\calD^*_ke^{-ikx}\right\},\label{167}
\ee
from which it follows that
\be
-\psidag_\E(x)\g^5_\E
=-\int \frac{d^4k}{\sqrt{(2\pi)^4}}\left\{
\calB_k^\dagger\g^5_\E 
e^{-ikx}+\calD^\top_k\g^5_\E e^{ikx}\right\},
\ee
where we have replaced the combinations 
$U_k\circ\calB_k$ and $D_k^\dagger\circ V_{-k}$ by the
operator-valued four-spinors $\calB_k$ and $\calD_k^*$,
respectively.
We still require that $\calB_k|0\rangle_\E
=0=\calD_k|0\rangle_\E$.

By virtue of the rotation on spinor indices, 
the unitary involution $\Theta$ is now defined on spinors in the
same way as for bosons
\bea
\Theta\psi_\E(x)\Theta\inv=\psi_\E(\Theta x)\ ; \
\Theta\psidag_\E(x)\Theta\inv=\psidag_\E(\Theta x),
\label{mirth}
\eea 
which in turn defines the action of $\Theta$ on the spinor modes
\bea
\Theta\calB^\dagger_k\Theta\inv&=&\calB^\dagger_{\Theta k},
\label{trieste}\\
\Theta\calD^\top_k\Theta\inv&=&\calD^\top_{\Theta k}.
\eea
Clearly this action is involutive.
Euclidean ultralocality,
\be
0=\{\psi_\E(x), 
-\psi_\E(y)^\dagger\g_\E^5\},\ee is satisfied by
requiring
\be
\{\calB_k,\calB^\dagger_{k^\prime}\}
=-\{\calD_{-k}^*,\calD_{-k^\prime}^\top\}.
\label{B1}
\ee
The correct two-point function as in\rf{2} is ensured
by the following anticommutation relations for the ``spinor
modes''
\be
-\{\calB_k,\calB^\dagger_{k^\prime}\g^5_\E\}=
\frac{-i\kslash^\E+m}{\O_k^2}
\delta^4(k-k^\prime).
\label{B2}
\ee
The matrix $\left(\frac{-i\kslash^\E+m}{\O_k^2}\right)(-\g_\E^5)$
is hermitean with eigenvalues $\pm\O_k$ and may be diagonalized
(in a basis for the Dirac matrices in which $\g_\E^5$ is
diagonal) by defining
\be
\calB_k=\frac{1}{\sqrt{\O_k}}\calU_k\widetilde{\calB}_k
\ee
where
\be
\calU_k=\frac{\O_k-m+i\kslash^\E}{\sqrt{2(\O_k-m)\O_k}}=
(\calU_k\inv)^\dagger=(\calU_{-k})^{-1}.
\ee
The combination $\frac{1}{\sqrt{\O_k}}\calU_k$ 
is the analogue
of the OS spinor wave function $U_k$. In this basis the mode
relations now read
\be
\{\widetilde{\calB}_k,\widetilde{\calB}^\dagger_{k^\prime}\}
=-\g^5_\E\delta^4(k-k^\prime).\label{B4}
\ee 
That the Euclidean Fock space now contains negative norm states
causes no problems since we only require positive norm in the
Minkowski Hilbert space, which is assured since, by construction,
our fields satisfy the same  axioms as those of the OS proposal.
(It is amusing to note that although we still double the number 
of
mode operators, we find that half of them have negative norm 
in the 
Euclidean Fock space).

We may ``diagonalize'' the $\calD_k$ modes in a similar 
fashion so that the final result for our Euclidean fields reads
\be
\psi_\E(x)=\int \frac{d^4k}{\sqrt{(2\pi)^4\O_k}}\left\{
U_k\widetilde{\calB}_ke^{ikx}+U_{-k}
\widetilde{\calD}^*_ke^{-ikx}\right\},
\ee
and 
\be
\chi^\dagger_\E(x)=-\psidag_\E(x)\g_\E^5=
-\Theta\inv\psidag_\E(\Theta x)\g^5_\E\Theta.
\ee
The modes satisfy commutation relations
\be
\{\widetilde{\calB}_k,\widetilde{\calB}_{k^\prime}^\dagger\}
=-\g^5_\E\delta^4(k-k^\prime)=-\{\widetilde{\calD}^*_{-k},
\widetilde{\calD}^\top_{-k^\prime}\},
\label{ul}
\ee 
By construction these fields anticommute at all points 
in Euclidean 
space
\be
\{\psi_\E(x),\chi^\dagger_\E(y)\}\equiv
-\{\psi_\E(x),\psi^\dagger_\E(y)\g^5_\E\}=0
\ee
and possess the desired two-point function,
\be
_\E\!\lvac\psi_\E(x)\psidag_\E(y)(-\g_\E^5)\vac_\E=
\int\frac{d^4k}{(2\pi)^4}\frac{-i\kslash^\E+m}{k^2_\E+m^2}
e^{ik_\E(x-y)_\E}.\label{22}
\ee
which is consistent with hermiticity.
We have now reproduced the building blocks ((\ref{22}), 
(\ref{ul}), (\ref{mirth}) and\rf{hendrix}) of 
the OS construction and the rest of their proposal may now be
inherited unaltered
except for the
replacement everywhere of the 
field $\chi^\dagger_\E$ by $-\psi^\dagger_\E\g^5_\E$
and the extra  
rotation of spinor indices performed by the new OS--Wick map.

\section{Conclusion}

In this note we have presented a 
generalization of the canonical work of Osterwalder and Schrader
for Dirac spinors which fuses their approach with the
Schwinger--Zumino~\cite{Schwinger}\cite{Zumino} approach in which
hermiticity is maintained although we still found it necessary
to double the set of Euclidean fermionic mode operators.
This was achieved by considering a new Wick rotation for
fermions under which spinor indices also rotate. We would also
suggest, that our generalization adds to the formal simplicity
of the OS construction.

Finally, one may wonder what happens in the case of Majorana
or Weyl spinors. In~\cite{us} we found that for Majorana
and Weyl spinors in four dimensions 
that the requirement of hermiticity must be dropped. The
extension of the work of OS to Majorana spinors was given by
Nicolai~\cite{Nicolai} who noted that although there existed no
consistent reality condition for Majorana spinors in
four dimensional Euclidean
space, one could nonetheless define a symplectic reality
condition on the mode operators $B_k$ and $D_k$ following which
the OS proposal may also be simply inherited. It is not a
difficult 
matter to apply the generalizaton we have given above also to
Nicolai's work, and although one cannot regain hermiticity,
the same formal algebraic simplifications as above occur.
One may then even study $N=1$ supersymmetric systems. As usual
real bose fields undergo complex supersymmetry
transformations  
in Euclidean space since the canonical supersymmetry charge 
$Q$ no longer satisfies any reality condition. 
Such considerations may also be formulated
in superspace
(see also~\cite{Schrader}).

\section{Acknowledgements.}

I am deeply indebted to Peter van Nieuwenhuizen for 
numerous detailed discussions and suggestions.

\end{document}